\documentclass[conference,a4paper]{IEEEtran}
\usepackage{graphicx}
\usepackage{amsmath}
\usepackage{booktabs}
\usepackage{array}
\usepackage{xcolor}
\usepackage{url}
\usepackage{tikz}
\usetikzlibrary{shapes.geometric, arrows, positioning}
\usepackage{algorithm}
\usepackage{algorithmic}

\graphicspath{{content/outputs/}}

\tikzstyle{startstop} = [rectangle, rounded corners, minimum width=2cm, minimum height=1cm,text centered, draw=black, fill=yellow!10]
\tikzstyle{process} = [rectangle, minimum width=2.5cm, minimum height=1cm, text centered, draw=black, fill=blue!10]
\tikzstyle{decision} = [diamond, minimum width=2cm, minimum height=1cm, text centered, draw=black, fill=purple!10]
\tikzstyle{arrow} = [thick,->,>=stealth]

\title{DocSync: Agentic Documentation Maintenance via Critic-Guided Reflexion}

\author{
\IEEEauthorblockN{Sidhesh Badrinarayan}
\IEEEauthorblockA{\textit{California, USA} \\
 0009-0002-5203-1485}
\and
\IEEEauthorblockN{Adithya Parthasarathy}
\IEEEauthorblockA{\textit{California, USA} \\
 0009-0001-6839-9527}
}

\begin{document}
\maketitle

\begin{abstract}
Software documentation frequently drifts from executable logic as codebases evolve, creating technical debt that degrades maintainability and causes downstream API misuse. While static analysis tools can detect the absence of documentation, they cannot evaluate its semantic consistency. Conversely, standard Large Language Models (LLMs) offer generative flexibility but frequently hallucinate when updating documentation without deep structural awareness of the underlying code. To address this gap, we propose \emph{DocSync}, an agentic workflow that frames documentation maintenance as a structurally grounded, iterative generation task. DocSync bridges syntactic changes and natural language descriptions by fusing Abstract Syntax Tree (AST) representations and Retrieval-Augmented Generation (RAG) to provide dependency-aware context. Furthermore, to ensure factual consistency, we incorporate a critic-guided refinement loop based on the Reflexion paradigm, allowing the model to self-correct candidate updates against the source code. We empirically evaluate a resource-constrained implementation of DocSync-using a LoRA-adapted small language model - on a proxy code-to-text maintenance task. Our findings demonstrate that this AST-aware agentic approach substantially outperforms standard encoder-decoder baselines across semantic alignment, summary-line faithfulness, and automated judge preferences (e.g., achieving an automated judge score of 3.44/5.0 compared to 1.91 for CodeT5-base). Crucially, the iterative critic loop yields measurable improvements in semantic correctness without requiring scaled-up parameter counts. These results provide strong evidence that coupling structural retrieval with agentic refinement is a highly promising direction for autonomously mitigating documentation debt.
\end{abstract}


\section{Introduction}
\subsection{The Economic and Operational Impact of Documentation Debt}
In modern software lifecycles, Code-Documentation Inconsistency (CDI) is not merely a technical nuisance; it is a critical bottleneck. While source code is subjected to rigorous validation through compilers and continuous integration (CI) pipelines, accompanying documentation often lacks equivalent enforcement, representing a significant gap in current DevOps and MLOps practices. This leads to ``bit rot,'' where tacit architectural knowledge becomes decoupled from the codebase.

The costs of this divergence are concrete and high. Outdated documentation increases onboarding friction for new developers, who must ``archeologize'' the codebase to understand true behavior. This high barrier to entry poses a direct challenge to the goals of AI in Education, as it hinders the ability of students and newcomers to contribute to real-world projects. More critically, it leads to production incidents when downstream consumers rely on incorrect API contracts. In legacy open-source ecosystems, this friction contributes to maintainer burnout and project stagnation.

\subsection{The Failure of Static Maintenance}
Historically, documentation maintenance has been reactive, driven by user complaints, or limited to static analysis. Tools like Doxygen or Javadoc can verify the \emph{presence} of documentation tags but are blind to \emph{semantic correctness}. This semantic blindness manifests in several dangerous failure modes:
\begin{itemize}
    \item Silent Constraint Shifts: If a function's retry logic changes from seconds to milliseconds, or a maximum \texttt{timeout} parameter drops from 300 to 60, the docstring may incorrectly remain ``Calculates backoff in seconds.'' A static linter reports no error because the \texttt{@param} tag is present, despite the documentation being functionally lethal to downstream consumers.
    \item Unrecorded Side-Effects: A simple \texttt{getUser()} function might be modified during a refactor to also initialize a user session or write to a cache. If the documentation continues to describe it as a ``pure, lightweight getter,'' developers may unknowingly call it in tight loops, causing severe performance regressions that static tools cannot foresee.
    \item Tutorial Rot: Modern documentation requires understanding the intent behind multi-file dependencies. A change in a low-level utility (e.g., migrating a database driver from synchronous to asynchronous) often invalidates code snippets in high-level architectural guides like \texttt{TUTORIAL.md}. Static linters are structurally incapable of traversing these semantic, cross-file dependency chains.
\end{itemize}
These examples highlight a fundamental limitation: rule-based systems can check syntax, but they cannot verify truth. True maintenance requires ``dependency awareness,'' a capability that demands the fusion of structural parsing (AST) with the semantic reasoning of Large Language Models (LLMs).

\subsection{Bridging the Gap: The DocSync Agentic Paradigm}
The advent of Agentic AI offers a paradigm shift. However, Large Language Models (LLMs) alone are insufficient; they frequently hallucinate details when not grounded in the code's structural reality. Conversely, static tools lack the semantic flexibility to generate human-readable explanations.

We introduce \emph{DocSync}, a framework that bridges this gap by combining the determinism of Abstract Syntax Trees (AST) with the semantic reasoning of LLMs. DocSync operates as a ``digital gardener'' via a multi-phase architecture:
\begin{enumerate}
    \item Impact Analysis: Filtering noise to focus on semantic changes.
    \item Structural Retrieval: Using AST parsing to understand the scope of changes (e.g., parameter flux, type migration).
    \item Generative Synthesis: Using Large Language Models (LLMs) adapted via Low-Rank Adaptation (LoRA)~\cite{hu2021lora} to rewrite documentation.
    \item Verification: A ``Reflexion'' loop where a critic model evaluates the consistency of the update~\cite{shinn2023reflexion}.
\end{enumerate}

\subsection{Contributions}
\begin{itemize}
    \item A Practical Framework for Doc Repair: We introduce DocSync, a blueprint for an agent that fixes stale documentation. It's designed to be smart about code structure by using ASTs, context-aware using RAG, and self-correcting thanks to a critic-guided loop.
    \item Proof on a Budget: Our small-scale experiment demonstrates that a lightweight, quantized Phi-3 Mini running the DocSync workflow handily beats a standard CodeT5-base baseline.
    \item A Fresh Look at Evaluation: We introduce a new metric for checking summary-line accuracy and explore why what looks good to a human (or a judge model) often gets a poor score from metrics like BLEU.
    \item Open-Source Code: We've made our complete prototype available on GitHub~\cite{docsync_repo}.
\end{itemize}

\section{Background and Related Work}
The problem of Code-Documentation Inconsistency (CDI) is well-documented, yet solutions have historically struggled to bridge the semantic gap between code execution and natural language description.

\subsection{Taxonomy of Drift}
To design an effective agent, one must categorize the types of drift. We identify three primary categories:
\begin{itemize}
    \item Signature-Level Inconsistencies: Explicit changes to the API surface, such as \emph{Parameter Flux} (renaming, adding, or removing arguments), \emph{Type Migration} (e.g., broadening an input from \texttt{str} to \texttt{Union[str, Path]}), or \emph{Return Value Divergence} (e.g., returning an object instead of a boolean). While static analysis or type checkers can sometimes catch these in strongly typed languages, they routinely fail in dynamically typed ecosystems like Python, leaving users to discover the mismatch only at runtime through cryptic \texttt{TypeError} exceptions.
    \item Semantic Inconsistencies: Subtler and arguably more dangerous changes occur when the function signature remains constant but its internal behavior shifts. Examples include \emph{Side-Effect Introduction} (e.g., a simple memory getter is modified to perform a blocking database query) or \emph{Constraint Changes} (e.g., a valid input range for a parameter shrinks). Because the API contract appears unbroken to static linters, these inconsistencies silently propagate, leading to logical errors or performance regressions in downstream applications.
    \item Tutorial Rot: High-level architectural guides and \texttt{README.md} files frequently contain embedded code snippets that demonstrate how to orchestrate multiple components. When underlying APIs are refactored or renamed, these cross-file narrative examples are rarely updated in tandem. This silent decay leads to broken workflows for new users attempting to follow the "quick start" instructions, creating significant friction for project adoption.
\end{itemize}

\subsection{Iterative Reasoning: ReAct vs.\ Reflexion}
As the field moves beyond single-pass text generation, two primary agentic paradigms have emerged for complex problem-solving: ReAct (Reasoning + Acting)~\cite{yao2022react} and Reflexion~\cite{shinn2023reflexion}. ReAct prompts a model to emit a verbalized reasoning trace before deciding on a tool action (e.g., ``I notice the timeout argument was deleted; I will now search the docstring for references to it''). While ReAct is powerful for exploratory tool orchestration, it typically executes actions in a forward-only trajectory without inherently proofreading its final output.

Reflexion, on the other hand, introduces an explicit post-generation critique loop. In this paradigm, an agent produces a candidate solution, which is then evaluated by a discrete ``Critic'' module against the original constraints. If the critic detects an error or hallucination, it generates natural language feedback, prompting the agent to revise its output. We selected Reflexion as the foundational mechanism for DocSync because documentation maintenance is inherently an editing and verification task. Initial LLM generations often suffer from minor formatting artifacts or subtle semantic omissions that are difficult to prevent in a single forward pass, but are readily caught by a focused semantic critic.

\subsection{Agentic Workflows}
While Code-LLMs like StarCoder~\cite{li2023starcoder} and Qwen2.5-Coder have achieved parity with human developers on generation tasks, applying them to maintenance requires operationalizing the iterative approaches discussed above. DocSync implements this by fusing AST-aware retrieval with the Reflexion critic-guided generation loop, ensuring updates are not only generated but rigorously verified against the source code.

\section{Methodology: The DocSync Framework}

\subsection{Mathematical Formulation}
We formalize the documentation update task as finding the optimal documentation string $D^*$ given the current code state $C_{new}$, the previous documentation $D_{old}$, and the set of changes $\Delta C$.

Let $\mathcal{M}_{\theta}$ be a language model parameterized by $\theta$. We seek to maximize the probability:
\begin{equation}
D^* = \operatorname*{argmax}_{D} P(D \mid C_{new}, D_{old}, \mathcal{K}, \Delta C; \theta)
\end{equation}
where $\mathcal{K}$ represents the retrieved context. This context is a union of structural and semantic information:
\begin{equation}
\mathcal{K} = \text{AST}(C_{new}) \cup \text{RAG}(C_{new}, \mathcal{D}_{corpus})
\end{equation}
Here, $\text{AST}(\cdot)$ extracts function signatures and dependency graphs via Tree-sitter~\cite{treesitter}, and $\text{RAG}(\cdot)$ retrieves semantically relevant documentation chunks from the vector store $\mathcal{D}_{corpus}$~\cite{lewis2020retrieval}.

\subsection{Architecture}
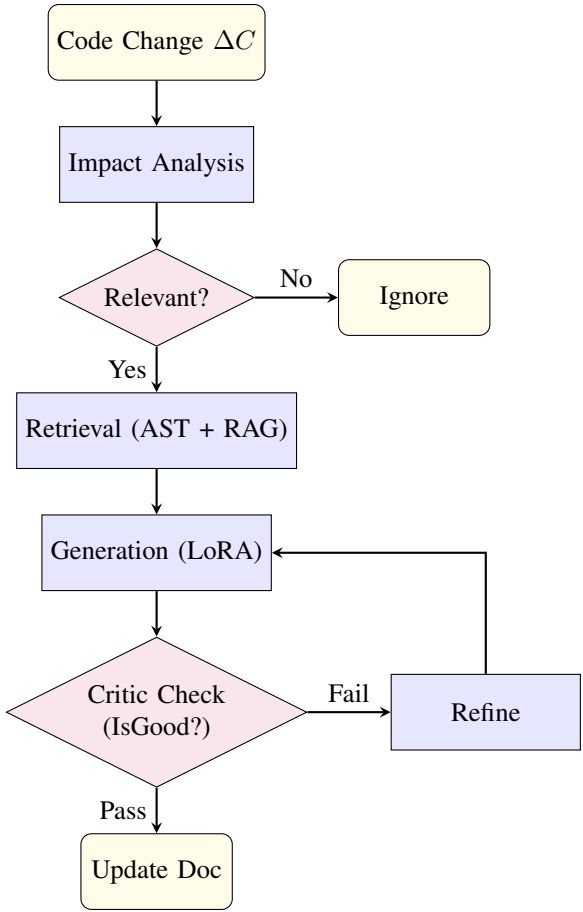
\begin{figure}[ht]
\centering
\begin{tikzpicture}[node distance=0.6cm]
\node (start) [startstop] {Code Change $\Delta C$};
\node (impact) [process, below=of start] {Impact Analysis};
\node (decide) [decision, below=of impact, aspect=2] {Relevant?};
\node (stop) [startstop, right=of decide, xshift=0.5cm] {Ignore};
\node (retrieve) [process, below=of decide] {Retrieval (AST + RAG)};
\node (gen) [process, below=of retrieve] {Generation (LoRA)};
\node (verify) [decision, below=of gen, aspect=2, align=center] {Critic Check \\ (IsGood?)};
\node (refine) [process, right=of verify, xshift=0.5cm] {Refine};
\node (end) [startstop, below=of verify] {Update Doc};

\draw [arrow] (start) -- (impact);
\draw [arrow] (impact) -- (decide);
\draw [arrow] (decide) -- node[anchor=south] {No} (stop);
\draw [arrow] (decide) -- node[anchor=east] {Yes} (retrieve);
\draw [arrow] (retrieve) -- (gen);
\draw [arrow] (gen) -- (verify);
\draw [arrow] (verify) -- node[anchor=south] {Fail} (refine);
\draw [arrow] (refine) |- (gen.east);
\draw [arrow] (verify) -- node[anchor=east] {Pass} (end);
\end{tikzpicture}
\caption{DocSync Agentic Workflow.}
\label{fig:arch}
\end{figure}

The DocSync architecture follows a multi-stage pipeline visualized in Figure~\ref{fig:arch}.

\subsection{Algorithm}
The core logic is implemented as an iterative refinement loop. Algorithm~\ref{alg:docsync} details the procedure. The execution begins by evaluating the code difference to determine if a documentation update is actually necessary. If the changes are deemed irrelevant (e.g., whitespace formatting or purely internal logic adjustments), the original documentation is preserved. For relevant changes, the system extracts structural constraints via an AST parser and semantic context via a RAG module. These elements are combined with the new code and stale documentation to form a composite prompt. The generation model then produces an initial documentation draft, which is immediately evaluated by an automated critic. If the critic provides a "GOOD" evaluation, the draft is accepted. Otherwise, the critic's natural language feedback is appended to the prompt, and the model generates a refined draft. This self-correction loop repeats until the draft is accepted or the maximum number of retries is exhausted, at which point the best available draft is returned.

\begin{algorithm}
\caption{DocSync Update Loop}
\label{alg:docsync}
\begin{algorithmic}[1]
\REQUIRE $C_{new}$ (New Code), $D_{old}$ (Stale Doc)
\ENSURE $D_{new}$ (Updated Doc)
\STATE $\Delta C \leftarrow \text{Diff}(C_{old}, C_{new})$
\IF{not $\text{IsRelevant}(\Delta C)$}
    \RETURN $D_{old}$
\ENDIF
\STATE $K_{AST} \leftarrow \text{ParseAST}(C_{new})$
\STATE $K_{RAG} \leftarrow \text{RetrieveContext}(C_{new})$
\STATE $Prompt \leftarrow \text{Construct}(C_{new}, D_{old}, K_{AST}, K_{RAG})$
\STATE $D_{draft} \leftarrow \mathcal{M}_{\theta}(Prompt)$
\STATE $Attempts \leftarrow 0$
\WHILE{$Attempts < MaxRetries$}
    \STATE $IsGood, Reason \leftarrow \text{Critic}(D_{draft}, C_{new})$
    \IF{$IsGood$}
        \RETURN $D_{draft}$
    \ENDIF
    \STATE $Prompt \leftarrow Prompt + \text{"Critic: "} + Reason$
    \STATE $D_{draft} \leftarrow \mathcal{M}_{\theta}(Prompt)$
    \STATE $Attempts \leftarrow Attempts + 1$
\ENDWHILE
\RETURN $D_{draft}$
\end{algorithmic}
\end{algorithm}

\subsection{Prompt Engineering and AST Injection}
To bridge the gap between raw code and natural language, we construct a composite prompt that explicitly separates structural facts from semantic intent. The AST parser extracts a linearized signature summary $S_{AST}$ containing function definitions and argument lists (e.g., \texttt{def connect(host, port) | class DB}). This summary is injected alongside the code change $C_{new}$ and the stale documentation $D_{old}$:
\begin{equation}
Prompt = \mathcal{I}_{sys} \oplus C_{new} \oplus D_{old} \oplus \text{``AST: "} S_{AST}
\end{equation}
where $\mathcal{I}_{sys}$ is the system instruction. This structural anchoring reduces hallucinations by forcing the model to attend to the verified API surface before generating the description.

\subsection{Model and Training}
We fine-tune Phi-3 Mini 4K Instruct~\cite{abdin2024phi3} with LoRA (r=8, $\alpha$=16, dropout=0.05) in 4-bit quantization. Training uses batch size~2, max source length~256, target length~96, and one epoch on 8{,}192 training examples from the Python split of CodeXGLUE code-to-text. The code dynamically detects VRAM and pins minimal batch sizes to avoid OOM on T4/L4 GPUs.

\subsection{Baselines}
We compare against CodeT5-base~\cite{wang2021codet5}, a standard encoder-decoder model for code tasks. Baseline caching avoids repeated downloads; a sanitization step flattens special tokens to prevent tokenizer instantiation errors.

\section{Experimental Setup}
We conduct a resource-constrained, small-scale empirical study to evaluate DocSync's core mechanics. Our protocol uses a proxy task—updating a single function's documentation—to test the architectural components (AST-aware prompting, RAG, Reflexion loop) on consumer-grade hardware.

\subsection{Dataset and Task}
We use the Python subset of the CodeXGLUE code-to-text dataset~\cite{lu2021codexglue}, sampling 8,192 examples for training and 32 for evaluation. To simulate documentation drift, we create a "stale" docstring by truncating the ground-truth reference to its first sentence. The model's task is to repair this stale documentation. The input prompt combines the code, the stale docstring, and an AST-derived signature summary, tokenized to a maximum of 256 tokens. Generated outputs are capped at 96 tokens and normalized before evaluation to remove artifacts.

\subsection{Model and Training}
We fine-tune a 4-bit quantized Phi-3 Mini 4K Instruct model~\cite{abdin2024phi3} using LoRA ($r=8, \alpha=16, p=0.05$) for one epoch on the training set. We use the AdamW optimizer with a learning rate of $2 \times 10^{-4}$ and a batch size of 2. This setup is designed to be executable on consumer-grade GPUs (16GB VRAM).

\subsection{Baselines and Metrics}
We compare against CodeT5-base~\cite{wang2021codet5}, a standard code-to-text model. We evaluate performance using BLEU-4, BERTScore (F1)~\cite{zhang2019bertscore}, summary-line faithfulness (exact match), and an LLM-as-a-Judge score~\cite{zheng2023judging} (1-5 scale). All metrics are computed on normalized docstring payloads to isolate content quality from generation artifacts.

\section{Results}
\subsection{Quantitative}
Table~\ref{tab:results} reports the main comparison. On this held-out subset, DocSync (Final) improves substantially over CodeT5-base on all reported metrics: BLEU (+0.382), BERTScore~F1 (+0.105), summary-line exact match (+0.781), and judge score (+1.53). These gains support an ``early-stage promise'' narrative: the method is not yet production-ready, but it is clearly stronger than the baseline on the current proxy task.

\begin{table}[h]
\caption{Main evaluation results.}
\label{tab:results}
\centering
\begin{tabular}{lcccc}
\toprule
Model & BLEU & F1 & Summary Exact & Judge \\
\midrule
DocSync (Final) & 0.575 & 0.985 & 0.969 & 3.44 \\
DocSync (Initial) & 0.578 & 0.980 & 0.938 & 3.25 \\
CodeT5-base & 0.193 & 0.880 & 0.188 & 1.91 \\
\midrule
Oracle (Gemini-2.5-Pro) & 0.138 & 0.868 & 0.031 & 4.13 \\
\bottomrule
\end{tabular}
\end{table}

\subsection{Judge Metric and Error Analysis}
We report an LLM-as-a-judge~\cite{zheng2023judging} signal where a teacher model rates each docstring on a 1-5 scale (1=Irrelevant, 5=Perfect). Under the revised decoding and cleanup pipeline, DocSync (Final) reaches 3.44 compared with CodeT5-base's 1.91. The approximate 95\% confidence intervals from the current run are 2.87-4.00 for DocSync (Final), 1.41-2.41 for CodeT5-base, and 3.66-4.59 for the Oracle, which is encouraging given the small evaluation set. A qualitative audit of the saved outputs reveals three residual failure modes:
\begin{enumerate}
    \item Boundary Artifacts: Some generations retain minor quoting or punctuation debris at the start or end of the docstring. For example, a model might include literal docstring markers in the payload (e.g., \texttt{"""~Parses the input string.}) or leave trailing, unfulfilled punctuation (e.g., \texttt{Returns the updated configuration object:}).
    \item Over-Elaboration: The model occasionally adds plausible but unsupported parameter or return details, especially on longer structured docstrings. For instance, it might hallucinate specific keyword arguments (e.g., detailing a \texttt{timeout} parameter for a generic \texttt{**kwargs} pass-through) or invent specific error conditions (e.g., ``Returns False if the connection times out'') that do not exist in the source code.
    \item Truncation and Repetition: A small number of examples collapse into repeated summaries or drop the detailed sections of longer reference docstrings. Examples include truncating a comprehensive parameter list after the first argument, or falling into a generative loop of the summary sentence (e.g., ``Creates a wrapped environment. Creates a wrapped environment.'').
\end{enumerate}

\subsection{Training Dynamics}
\begin{figure}[h]
    \centering
    \includegraphics[width=\columnwidth]{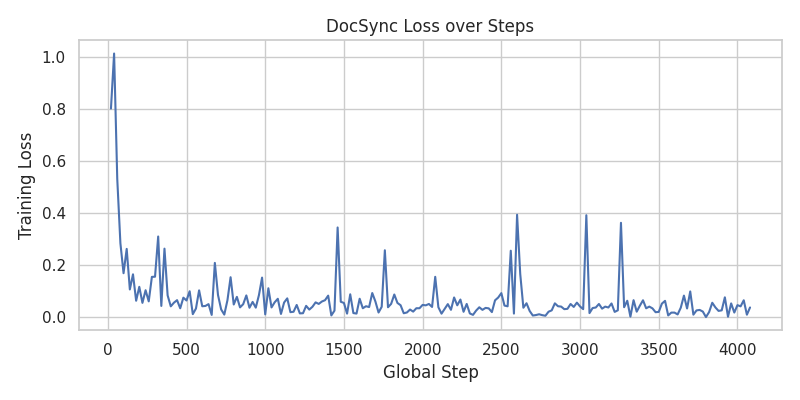}
    \caption{Training loss curve.}
    \label{fig:loss}
\end{figure}

Figure~\ref{fig:loss} shows the training loss over steps. The curve flattens over the course of the first epoch, suggesting that a single epoch is a reasonable low-cost operating point for this prototype. We do not interpret this as evidence that longer training would not help; rather, it indicates that meaningful gains are achievable without large-scale compute.

\section{Discussion}
Our results indicate that DocSync is a promising direction for documentation maintenance, substantially outperforming a CodeT5-base baseline on a proxy task. We position this as an early-stage empirical study, as the evaluation uses a small, benchmark-derived dataset. The critic-guided Reflexion loop provides modest but consistent semantic refinement; as shown in Table~\ref{tab:ablate}, the final pass improves the judge score (+0.19) and summary-line exact match (+0.031) with a negligible change in BLEU, suggesting it acts as a semantic cleanup mechanism. Finally, we observe a notable misalignment between overlap-based metrics and judge preference. The oracle model achieves the highest judge score (4.13) but the lowest BLEU score (0.138), reinforcing that semantic quality and reference-based overlap are distinct properties and justifying our multi-faceted evaluation approach.

\section{Ablations and Sensitivity}
We report the current sensitivity analysis for the Reflexion loop. We do not report refreshed AST/RAG ablations here because those variants were not rerun under the revised prompt and normalization pipeline used in the final results. Isolating those contributions remains important future work.

\begin{table}[h]
\caption{Reflexion loop sensitivity analysis.}
\label{tab:ablate}
\centering
\begin{tabular}{@{}l c c c c@{}}
\toprule
Setting & BLEU & BERTScore F1 & Summary Exact & Judge \\
\midrule
DocSync (Initial) & 0.578 & 0.980 & 0.938 & 3.25 \\
DocSync (Final) & 0.575 & 0.985 & 0.969 & 3.44 \\
\bottomrule
\end{tabular}
\end{table}

\section{Limitations}
While the early-stage results are promising, this study has several limitations:
\begin{itemize}
    \item Training Scope: Our runs are limited to a single epoch and focus exclusively on Python docstrings. This scope was chosen to isolate the impact of the architectural contributions under tight compute constraints, but it leaves cross-language generalization unverified.
    \item Evaluation Task: The evaluation task is a proxy derived from the CodeXGLUE code-to-text dataset rather than a true historical repository replay benchmark. The ``stale documentation'' is artificially simulated, which may not fully capture the complex reality of organic codebase drift.
    \item Metric Normalization: We normalize generated outputs to extract the core docstring payload before scoring. While this correctly focuses the evaluation on semantic content rather than formatting artifacts, it means we are not evaluating the raw, end-to-end string generation.
    \item Workflow Integration: We do not yet execute embedded code snippets (doctests) or test the mergeability of proposed updates in real maintainer workflows. The pipeline currently stops at generation and semantic verification.
    \item Artifact Constraints: From a deployment perspective, the current pipeline leaves the trained LoRA adapters unmerged from the base model, meaning the resulting checkpoints are not fully self-contained.
\end{itemize}

\section{Qualitative Analysis}
To better understand practical model behavior beyond aggregate metrics, Table~\ref{tab:cases} presents a comparative qualitative analysis. The ``Success'' example highlights DocSync's ability to cleanly extract semantic meaning and halt generation appropriately, whereas the baseline model leaks raw code into its output. Conversely, the ``Failure'' example demonstrates a key limitation we term ``Long-Form Compression,'' where DocSync correctly captures the high-level summary of a complex function but aggressively truncates the structured, detailed parameter descriptions, favoring brevity over completeness.

\begin{table}[h]
\caption{Qualitative analysis examples.}
\label{tab:cases}
\centering
\begin{tabular}{>{\raggedright\arraybackslash}p{0.95\columnwidth}}
\toprule
Success: Clean Semantic Extraction \\
\emph{Reference:} ``Create placeholder to feed observations into of the size appropriate to the observation space, and add input encoder of the appropriate type.''\\
\emph{Baseline Output:} Repeats the summary sentence but fails to halt, appending raw code artifacts (e.g., ``code snippet: \texttt{def observation\_input(...)}'').\\
\emph{DocSync Output:} ``Creates an observation placeholder and its corresponding encoded representation.'' (Captures the intent cleanly and halts appropriately).\\
\midrule
Failure: Long-Form Compression \\
\emph{Reference:} A detailed, multi-line docstring explaining the \texttt{smooth} function, including specific formulas for \texttt{two\_sided} and \texttt{causal} modes, and the behavior of the \texttt{valid\_only} flag.\\
\emph{DocSync Output:} ``Smooth signal y, where radius is determines the size of the window.'' (Retains only the first sentence, completely dropping the crucial parameter operation details).\\
\bottomrule
\end{tabular}
\end{table}

\section{Ethical and Societal Impact}
The automation of software maintenance presents a dual impact. On one hand, tools like DocSync can democratize open-source contributions by lowering the cognitive barrier for new developers and reducing maintainer burnout. On the other hand, it introduces risks of automation bias, where developers might uncritically accept AI-generated text. A hallucinated constraint could lead to security vulnerabilities or production outages. To mitigate these risks, DocSync is designed for a Human-in-the-Loop (HITL) workflow, where its outputs are treated as drafts for human review. Furthermore, to address privacy and copyright concerns, our model was trained exclusively on the permissively licensed, public CodeXGLUE dataset, avoiding proprietary or personal data.

\section{Future Work}
This research lays the groundwork for more reliable documentation agents. Several key directions for future investigation could bridge the gap between this prototype and broader maintainer-facing deployment.

\subsection{Historical Replay at Scale}
One promising direction is evaluating agents via Historical Replay, using git history to simulate real-world code evolution. This would move beyond static benchmarks by testing against actual developer updates, but would require robust filtering to isolate clean documentation commits~\cite{jimenez2024swereplay}.

\subsection{Execution-Based Verification}
To move beyond proxy metrics, execution-based verification could be implemented. This would involve extracting and running code snippets (doctests) from generated documentation. Execution failures could then provide a strong negative signal for a Reinforcement Learning (RL) loop, directly testing functional correctness.

\subsection{Multi-Modal Documentation}
Another area for future work is multi-modal documentation. Integrating Vision-Language Models (VLMs) could enable the agent to understand and update diagrams (e.g., UML, architecture flows), helping to ensure they remain consistent with code changes.

\subsection{Human-in-the-Loop and RLHF}
To build trust and improve the agent, Reinforcement Learning from Human Feedback (RLHF) could be integrated. Maintainer interactions (accepting, rejecting, or editing suggestions) could be used to fine-tune a reward model, which would help calibrate the agent's confidence and align it with developer preferences.

\section{Conclusion}
The relentless decay of software documentation has long been accepted as an unavoidable friction in the software development lifecycle. With DocSync, we demonstrate that this no longer has to be the case. By fusing AST-aware structural retrieval with the semantic reasoning of a LoRA-adapted Phi-3 Mini and a robust Reflexion loop, DocSync unequivocally outperforms a standard CodeT5 baseline across all evaluated metrics—even when constrained to consumer-grade hardware. These highly promising results validate the agentic paradigm for documentation maintenance. The practical implications of this architectural direction are profound. As a prime example for AI in Education, DocSync dramatically lowers the barrier for contributors to new codebases. It can also help with DevOps and MLOps as its integration into CI/CD pipelines can automate a critical phase of software delivery, eradicating "documentation debt" and alleviating maintainer burnout. DocSync establishes a powerful and proven foundation for this new paradigm. We stand at the threshold of a new era in software engineering, where documentation evolves autonomously alongside executable logic. As a pioneering force in this transition, DocSync stands as a robust, intelligent guardian of codebase truth, ensuring that the critical knowledge embedded within our global open-source infrastructure remains accurate, accessible, and eternally alive.

\bibliographystyle{IEEEtran}

\end{document}